\documentclass[%
 reprint,
 amsmath,amssymb,
 aps,
]{revtex4-2}

\usepackage{graphicx}
\usepackage{dcolumn}
\usepackage{bm}
\usepackage{float}
\usepackage{comment}
\usepackage{subfigure}
\usepackage{multirow}

\begin{document}

\preprint{APS/123-QED}

\title{Cosmic Ray Induced Neutron Production in a Lead Target}
\author{Haichuan Cao}
\author{David Koltick}
\affiliation{Department of Physics and Astronomy\\ 525 Northwestern Ave., West Lafayette, Indiana USA, 47907}




\date{\today}


\begin{abstract}
\noindent
Underground experiments searching for rare events, such as interactions from dark matter, need to exhibit background as low as possible. One source of background is from cosmic ray muons and muon-induced neutron production.   Presently these background are not fully understood. In this study Geant4 is used to model cosmic ray muon induced neutron multiplicity production and compare the modeling with data collected using an $^3$He instrumented Pb-target detector system. 
The neutron event multiplicity production is taken from the 2002 NMDS-II data sets, consisting of 6504  hrs collected at 583 m.w.e. and  1440 hrs, with the identical detector system, collected at 1166 m.w.e.. The detector consists of a 30 cm cube Pb-target surrounded by 60 $^3$He tubes.  The single particle detection efficiency is 23.2\%$\pm$1.2\%  calibrated using a $^{252}$Cf neutron source.  The highest neutron multiplicity event, observed at 583 m.w.e. was 54 observed neutrons corresponding to $\sim$ 233 produced neutrons.  
The neutron multiplicity, n, distributions  fit well a 2-parameter power law fit, $k\times n^{-p}$. For the Monte Carlo simulations at both depths and the data collected at both depths, all are consistent with a single slope parameter p. For the simulation at 583 m.w.e., p=2.37$\pm0.01$ and for the data collected at 583 m.w.e, p=2.36$\pm0.10$. At 1166 m.w.e., p=2.31$\pm0.01$ for the simulation, and for the data with only 6 detected events above multiplicity 5, p=$2.50 \pm 0.35$ predicted using a Maximum Likelihood Estimation method.  At both depths, the power law amplitudes of the Geant4 simulations differ by a factor of 2 larger than the data sets.  However, the disagreement is within the estimated systematic error of the simulations.

\end{abstract}

\maketitle


\section{Introduction}
Experiments searching for rare events, such as interactions from dark matter, need to exhibit radioactive backgrounds as low as possible. Methods to reduce such backgrounds include the selection of radio-pure materials for building components \cite{kudryavtsev2008neutron}, and placing the experiment deep-underground \cite{kudryavtsev2008neutron}.  Even so, some backgrounds are not avoidable, the most critical of which is background from muon-induced neutrons. These can either generate prompt signals in detectors or produce long lived radioactive isotopes by capture or inelastic scattering reactions. The subsequent decay of the  resulting isotopes can lead to significant backgrounds in the detectors if the correlation to the corresponding muon is lost \cite{best2016low}. To shield detectors from the radioactivity experimental components as well as from the surrounding environment,  high-Z materials, such as lead, are often selected.  
Being located close to the sensitive volume  such shielding is a source of muon-induced neutrons.

Furthermore, new physics processes initiated by cosmic ray or dark matter-matter interactions may be observable through excess production of high multiplicity neutron production in nuclear targets \cite{Ward2023}. For high energy particles, an elemental high mass number solid detector, can be thought of in a manner similar to a solid state  detector to $\gamma$-rays, except the band gap is of order $\sim$10 MeV. 

In both cases, whether to improve an understanding of radioactive backgrounds or as a new physics direct search method, further understanding of muon-induced neutron backgrounds is crucial in the search for BSM physics.

 To this end, we have analyzed the data collected by the NMDS-II detector, containing 305 kg of Pb,  at two depths, 583 m.w.e. (meters water equivalent) and 1166 m.w.e., located at the Center for Underground Physics in Pyhas\"almi(CUPP). The CUPP Underground Laboratory extends down to 1440 m (rock), within the Pyhas\"almi zinc, copper and pyrite mine complex in central Finland. Pyhas\"almi is the deepest metal mine in Europe\cite{enqvist2003infrastructure}.

At the experiment's depths neutrino to muon conversion interactions are $\sim10^{-6}$ below the muon rates\cite{crouch1978cosmic}.  The only source of multiple neutron production then is directly by muon induced production or by particles associated with muon induced showers  in the surrounding rock. To set the energy scale, a single $\sim$30 TeV muon is expected to pass through the detector in the collected data sets.  The data sets record single events producing up to $\sim$233 neutrons. 

For the 583 m.w.e. (220m rock) data, measurements were collected from February 28th 2002 to February 9th 2003. Taking into account stops for technical reasons, the live acquisition time was $6504 \pm 1$ hours. In this data set approximately 50\% of the events are due to a muon entering through the top and an equal fraction through the sides of the detector.   For the 1166 m.w.e. (440m rock) data set collected $1440 \pm 1$ hours live time, between September 13-th 2001 to January 28-th 2002. 

Reported here is a comparison of the two  NMDS-II  data sets with Geant4 simulations of cosmic ray muon induced neutron production, both directly by muons passing through the detector as well as by interactions in the detector by associated muon induce shower particles produced in the surrounding rock. Separate simulations comparing FLUKA and Geant4 neutron production for muons passing through lead is used to estimate the systematic error of the Geant4 neutron production simulations.

\section{The NMDS-II Experiment}

\subsection{Neutron Detection System}

\begin{figure}
	\centering
	\includegraphics[width=0.75\linewidth]{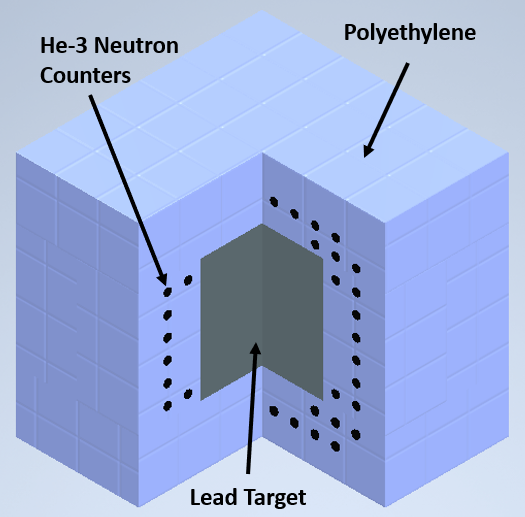}
	\caption{The NMDS-II detector, consists of a 30 cm cube Pb-target surrounded by 60, $^3$He detectors encased in polyethylene. The 60 cm polyethylene cube is composed of 4 side panels, one of which is displayed in cross section on the left side of the cut away.}
	\label{fig:CUPPdetector}
\end{figure}

The NMDS-II detector is an instrumented 30 cm cube of Pb (305 kg), which serves as the target. The Pb-cube is surrounded by a 60 cm cube polyethylene moderator 15 cm thick, within which 60, $^3$He proportional neutron counters, model SNM-18, produced by Maxiums Energy, are encased on all 6-sides, as shown in the Figure \ref{fig:CUPPdetector}. The $^3$He proportional tubes serve as the neutron multiplicity detection system. 

\begin{figure}
	\centering
	\includegraphics[width=0.8\linewidth]{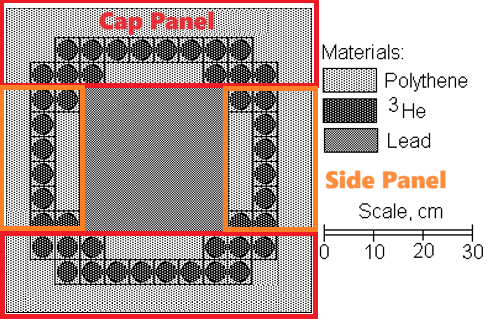}
	\caption{Cross section through the middle of the NMDS-II  detector showing the $^3$He neutron counters. The dimensions of the $^3$He tubes are not in proportion. The top and bottom cap panels are outlined in red.  The side panels are outlined in orange. }
	\label{fig:arrange_He3}
\end{figure}

 Each counter measures 28.5 cm long and 1.55 cm in diameter filled with a mixture of $^3$He (75\%) and Ar(25\%) at 4 atm pressure and operate in proportional mode at 1400 V. The neutron counters are arranged in 2 cap panels and 4 side panels forming the polyethylene box as shown in the Figure \ref{fig:arrange_He3}. The positions of the $^3$He counters were chosen by Monte-Carlo modeling, optimized so that neutron events occurring at any point in the target would be registered with almost the same efficiency.

A single neutron hit triggers the $^3$He counter system to record data for 256 $\mu s$. Each hit causes the stuck tube to be non-responsive or dead for a period of 10 $\mu s$. The neutron half-life in the lead target is 65 $\mu s$, resulting in an average of 6.47\% of the total neutrons produced to be outside the data collection window.

\begin{figure}
\centering

\begin{subfigure}
\centering
\includegraphics[width=0.9\linewidth]{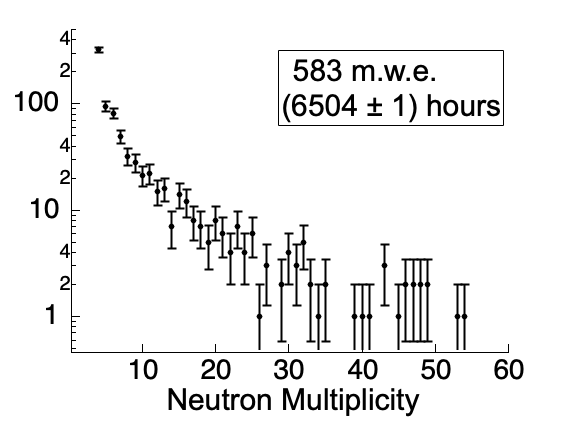}
\end{subfigure}
\begin{subfigure}
\centering
\includegraphics[width=0.9\linewidth]{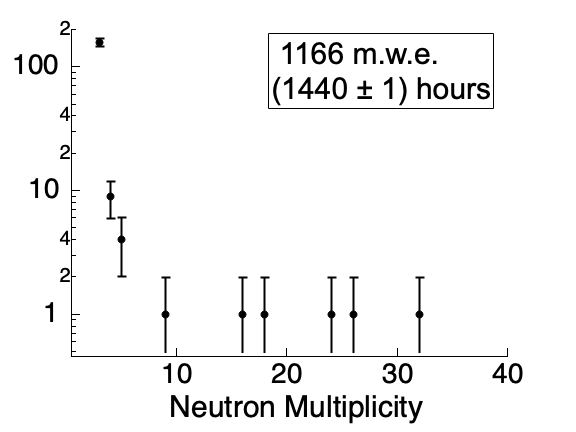}
\end{subfigure}

\caption{Observed cosmic ray muon induced neutron multiplicity distributions at 583 m.w.e. (Upper) and at 1166 m.w.e. (Lower), for the observation of 4 or more neutrons.}
\label{fig:CUPPdata}
\end{figure}

The two neutron multiplicity data sets collected by the NMDS-II, at 583 m.w.e. and at 1166 m.w.e. are shown in Figure \ref{fig:CUPPdata}.

\subsection{Experimental Analysis Event Sample}

To assure a pure sample of cosmic ray induced neutron events, a minimum of 4  observed neutrons are required to reduce the probability of neutron events produced by natural decays in the rock. At 1166 m.w.e., the observation rate of at least 1 neutron is $\sim 0.6\ n/s$, which includes natural decays and cosmic ray produced neutrons. This rate can be used to set an upper limit on the contribution of non-cosmic ray induced events populating the analysis sample.  The probability of 1 neutron in 256 $\mu s$, which is the event selection window, is $1.5 \times 10^{-4}$. If it is assumed that the number of decay neutrons follows a  Poisson distribution, the probability of 4 separate decay neutrons in a 256 $\mu s$ window is $2.3\times 10^{-17}$, and its contribution to the analysis sample is displayed in Table \ref{ta:natural_neutron}.

Because the single independent neutrons contribution is so low,  the $n \rightarrow 2n$ process, with threshold at $E_n \sim 7.5 MeV$, and the $n \rightarrow 3n$ process, at $E_n \sim 16 MeV$ are considered.  If it is assumed that neutrons from natural decays in the rock have an energy distribution like a $^{252}Cf$ source, $E^{1/2}\cdot exp{(-E/1.3)}$, where E is the neutron energy, the probability for one rock neutron causing an $n \rightarrow 2n$ process is $9.1\times 10^{-3}$ or an $n \rightarrow 3n$ process is $1.9\times 10^{-5}$. Again, upper limit contributions of these backgrounds to the cosmic ray sample are displayed in Table \ref{ta:natural_neutron}.  These estimates have not considered the neutron detection efficiency of the NMDS-II detector system.  Doing so would further reduce  the probability of observing 4 neutrons form other than cosmic ray sources, significantly.

\begin{table}
\caption{The upper limit probability of 4 observed neutrons induced by  non-cosmic ray sources in the target during the experiment.}

\centering
\begin{tabular}{|l|c|c|c|}
\hline
&256 $\mu s$  & 1440 hours & 6504 hours \\ \hline
4 natural neutron & $2.3\times10^{-17}$ & $4.7\times10^{-7}$     & $2.1\times10^{-6}$     \\ \hline
\begin{tabular}[c]{@{}l@{}}3 natural neutron,\\  one $n\rightarrow2n$\end{tabular}  & $5.2\times10^{-15}$ & $1.1\times10^{-4}$     & $4.8\times10^{-4}$    \\ \hline
\begin{tabular}[c]{@{}l@{}}2 natural neutron, \\ both $n\rightarrow2n$\end{tabular} & $9.9\times10^{-13}$ & $2.0\times10^{-2}$     & $9.0\times10^{-2}$     \\ \hline
\begin{tabular}[c]{@{}l@{}}2 natural neutron, \\ one $n\rightarrow3n$\end{tabular}  & $2.2\times10^{-13}$ & $4.4\times10^{-3}$     & $2.0\times10^{-2}$     \\ \hline
\end{tabular}
\label{ta:natural_neutron}
\end{table}

\section{Neutron Detector System Modeling}

The system-wide single neutron detection efficiency was calibrated using a $^{252}$Cf source  placed at the center of the target and found to be 23.2\%. However, the systems response to $^{252}$Cf is not the response to muon induced neutron production because of the difference in their energy spectra. To accurately estimate the detection efficiency of a produced neutron, whether produced in the lead target or surrounding rock, a Geant4 model detector was built matching the geometry of the lead target, the polyethylene thermalizer and the arrangement of $^3$He counters. 

The $^3$He tubes detect neutrons through the reaction 
\begin{equation}
   ^{3}He + ^{1}n(thermal) \rightarrow ^{1}H + ^{3}H + Q(764 keV)
    \label{equ:N_capture}
\end{equation}

 In the Geant4 simulation, when reaction \eqref{equ:N_capture} is observed, the corresponding counter is considered to generate a signal or neutron hit, without modeling the complex ion particle motion within the gas and walls of the tubes. If the diameter of the $^{3}$He tube in the simulation is chosen to be 1.55 cm as manufactured, the average neutron efficiency for the $^{252}$Cf spectrum is calculated to be $(28.0\pm 0.1) \%$, which is larger than the measured result. The difference between the simulation predicted efficiency and the measured efficiency is due to the wall effect\cite{KnollDetector}. The wall effect reduces the detection efficiency because either the proton or the triton or both strike the detector's cathode  or the non-sensitive volume at the counters' end, producing a reduced energy pulse, less than the Q in Equation \eqref{equ:N_capture}. If the energy deposition is below the set threshold, no hit is recorded. 
 
 The wall effect, that is  the reaction products, $^{3}$He(n,p)$^{3}$H occurring within one mean free path length away from the wall, and the set threshold are then important  parameters in calculating the system's neutron detection efficiency.  The mean free path length is  calculable analytically \cite{shalev1969wall}. Because the $^3$He tubes thresholds were set close to Q, the neutron detection efficiency is reduced by a geometric multiplying factor, 82\%, due entirely  to the Wall Effect.  Correcting the $^{252}$Cf Monte Carlo calculated calibration efficiency yields MC($^{252}$Cf)$\times$ Wall\ Effect = 0.28 $\times 0.82 \approx 0.23$, in agreement with the measured $^{252}$Cf  result. 
 
 Based on these results, the method used to calculate the neutron detection efficiency was to assume a reduced active ${^3}$He diameter set to 1.33 cm and record as a hit any observation of reaction \eqref{equ:N_capture}. In order to keep the amount of polyethylene the same, the diameter of the tube is kept at 1.55 cm, but the tube has an empty gap to account for the wall effect and threshold setting.

\begin{figure*}[t]
\centering
\begin{subfigure}
\centering
    \includegraphics[width=0.32\textwidth]{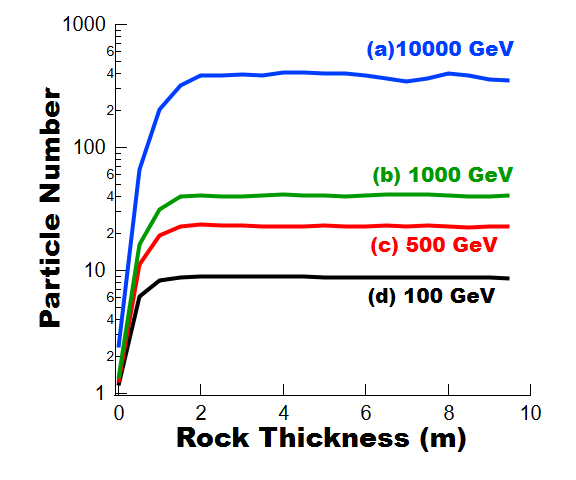}
\end{subfigure}
\hfill
\begin{subfigure}
    \centering
\includegraphics[width=0.32\textwidth]{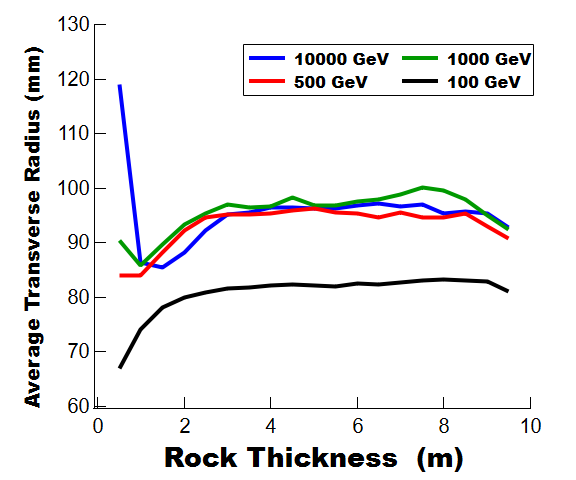}
\end{subfigure}
\hfill
\begin{subfigure} 
\centering
    \includegraphics[width=0.32\textwidth]{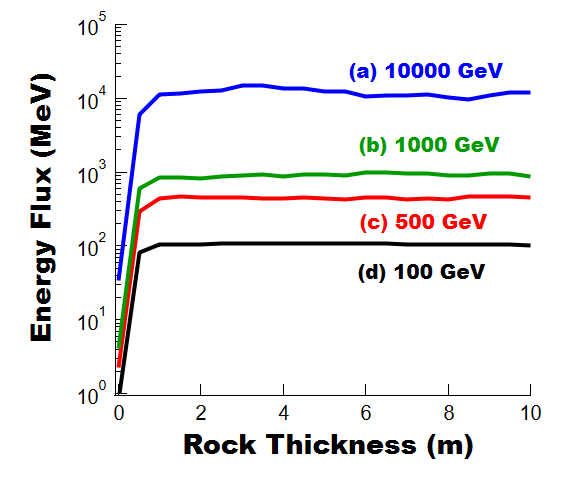}
\end{subfigure}
\caption{Simulated, averaged secondary particle flux (left), averaged secondary particle transverse radius (middle) and averaged secondary energy flux (right) passing through transverse planes as a function of muon path length.  The averaged secondary flux comes into equilibrium in $\sim$2 m independent of muon energy.}
\label{fig:Spectrum_Three}
\end{figure*}
\section{Muon Propagation to Depth}
\subsection{Muon Shower Equilibrium}

Cosmic ray muons are capable of penetrating to significant depths underground. At depth, these muon produce a myriad of particles, including photons, electrons, and hadrons. 
Other cosmic ray particles and those produced as the muon passes through the rock can travel only a short distance, $\sim$2m, in the rock. This short path length allows a simplification of the neutron production modeling because the full simulation does not need to start at sea level, but only a few meters above the  detectors laboratory hall.

To show this, muons with energy between 100 GeV to 10 TeV are propagated through rock having density 2.85g/cm$^2$ to find the path length after which the muon induced showers come into equilibrium. The reported observables are in the transverse plane to the muon's initial momentum as a function of the distance from the muon's entry into a 10 m thick rock layer.

Figure \ref{fig:Spectrum_Three} shows the results of the Geant4 simulation; (left) shows the averaged number of particles produced;  (middle) shows the averaged transverse radial component of 
secondary particle production. In this case the distribution is not zero at zero thickness. This effect occurs due to energy back-scatter from rock deeper along the muon's path. Figure \ref{fig:Spectrum_Three} (right)is the averaged secondary energy flux.  All the observables are evaluated on transverse planes to the muon's path. For all distributions, an equilibrium state is achieved within $\sim$ 2 m of rock, independent of muon energy. 

While secondary particle production quickly comes into equilibrium there is a concern that secondary muons may have a longer attenuation length than the equilibrium length of the dominating hadronic and electromagnetic components causing an additional muon flux on the Pb-target beyond the sea level component propagated to depth.  If a 2.5 GeV cutoff is used, equivalent to the minimum ionizing dE/dx energy loss of a muon passing through 4m of rock, it is found that secondary muons produced by primary muons with 100 GeV, 1 TeV, and 10 TeV, having a path length greater than 4m is less than 10$^{-5}$, 10$^{-4}$, 10$^{-3}$, allowing this secondary source of neutron production in the Pb-target to be neglected. 

From these distributions it is concluded to start the full muon shower simulation in a horizontal plane 4m above the detector's cavern hall captures the effects of secondary particle production due to the entire mines over burden, their contribution to neutron production outside the target, and their interactions with the lead target.

 \begin{figure}
\centering
\begin{subfigure}
\centering
\includegraphics[width=0.7\linewidth]{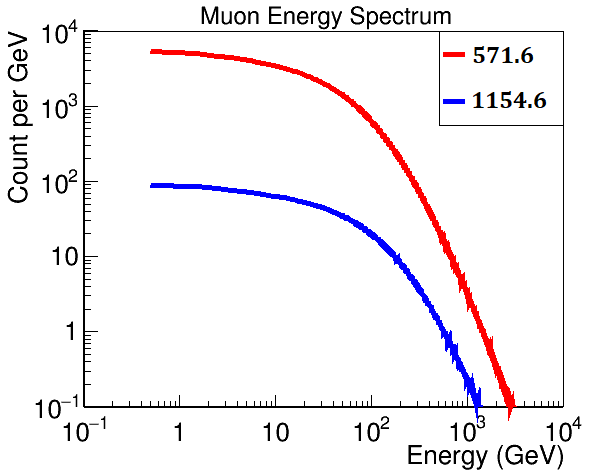}
\end{subfigure}

\begin{subfigure}
\centering
\includegraphics[width=0.7\linewidth]{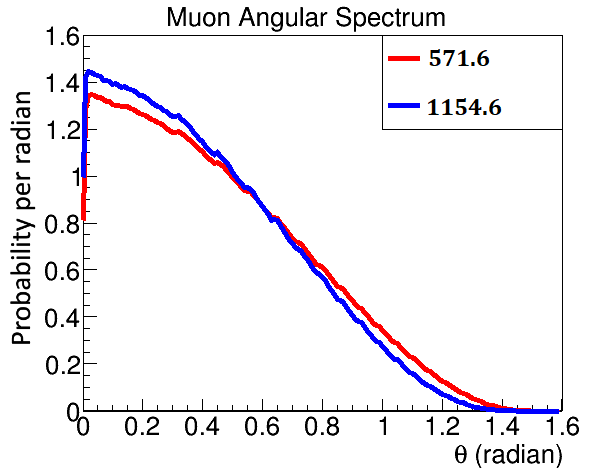}
\end{subfigure}

\caption{Cosmic ray muon energy and angular distributions at 571.6 mwe(6504 hr) and 1154.6 mwe(1440 hr). The energy spectra are normalized to the muon number passing through the target top during the data collection period. The mean muon energy for 571.6 m.w.e. is 95.6 GeV and for 1154.6 m.w.e. is 158.7 GeV. The total muon induced event number during the experiment at 571.6 m.w.e. is $\sim 36$ times that collected at 1154.6 m.w.e.. The two angular spectra are normalized to 1.}
\label{fig:underground_spectrum}
\end{figure}
\subsection{Muon Propagation from Sea Level to Depth}
To correctly model the correlated angular and energy distribution of muons at depth a propagation model must be used starting with the known correlated distributions at sea level.  No at depth experimental or analytical correlated distributions are available.  However, the validity of the Geant4 propagation model can be checked, using Miyake's empirical model to compare separately  the muon flux density and angular distributions as a function of depth. 

The correlated muon distributions at the two NMDS-II experimental depths are produced by starting with the well-measured energy-angular correlated analytic form of the sea-level cosmic ray muon spectrum \cite{CosmicRayPDG},

\begin{equation}
\begin{split}
    \frac{dN_\mu}{dE_\mu d\Omega} & \approx \frac{0.14E_\mu^{-2.7}}{cm^2\ s\ sr\ GeV}\times \\
&\left \{ \frac{1}{1+\frac{1.1E_\mu cos\theta}{115 GeV}} + \frac{0.054}{1+ \frac{1.1E_\mu cos \theta}{850 GeV}}\right \},
\end{split}
\label{equ:Surfcosmic}    
\end{equation}
 and propagating individual muons to depth.\

Formula \eqref{equ:Surfcosmic} is valid when muon decay is negligible ($E_\mu > $100/cos\hspace{0.05cm}$\theta$ GeV) and the curvature of the Earth can be neglected ($\theta < 70^\circ$). Fortunately, the two limitations do not influence the muon spectrum at 583 m.w.e and 1166 m.w.e underground.  As is discussed in Appendix \ref{app:muon_propogation}, muons not suitable for formula \eqref{equ:Surfcosmic} will not pass through the thick  583 m.w.e. rock layer.

\begin{figure}
\centering

\begin{subfigure}
\centering
\includegraphics[width=0.7\linewidth]{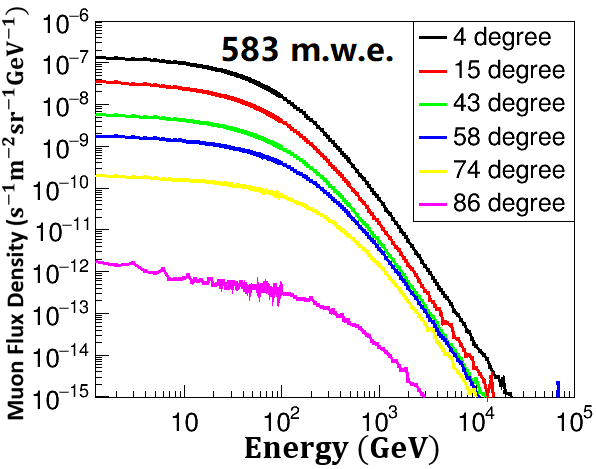}
\end{subfigure}
\begin{subfigure}
\centering
\includegraphics[width=0.7\linewidth]{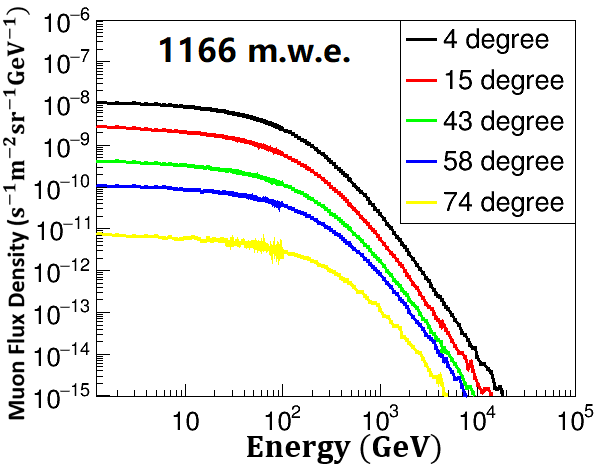}
\end{subfigure}

\caption{Muon flux density as a function of angle and incident energy at the experimental depths.}
\label{fig:Underground_spectrum_angle}
\end{figure}

 In the Geant4 model, muons are scattered in the rock, lose energy, create shower particles, and can suffer catastrophic energy loss. The muons are propagated from sea level to 4 m above the laboratory. Only muons are tracked to save simulation time. Shower particles are not tracked once generated. Finally, at depth, the muon's new energy (E'), and angles ($\theta '$ and $\phi '$) are recorded, having the required energy-angular correlation. The generated underground energy and angular spectra are shown in Figure \ref{fig:underground_spectrum}. To set the scale for illustration, these energy distributions have been normalized to the number of muons passing through the 30 cm x 30 cm top of the lead target, integrated over the observation time of each data set. Figure \ref{fig:Underground_spectrum_angle} shows the Geant4 calculated muon energy flux density at depth as a function of angle for the two locations of the detector 4m above the detector halls.

\subsection{Geant4 and Miyake Formula Comparison }  \label{sec:cosmic_miyake}

The underground intensity of muons can be calculated with the widely used empirical formula of Miyake.  At intermediate depths (100 m.w.e. to 4000 m.w.e.), the Miyake formula is given by \cite{kf2001cosmic} 

\begin{equation}
  I(X) = \frac{A}{X+400}(X+10)^{-1.53}e^{-8.0\times 10^{-4}X}, 
\label{equ:depth}
\end{equation}

\noindent where, \textbf{A} is the only free parameter and \textbf{X} is in m.w.e..  \textbf{A} was found by fitting measurements made at the Pyhas\"almi mine above and below the experiment\cite{enqvist2005measurements}. The flux density measurements are presented in Appendix \ref{app:CUPP_depth}. The fit returned $\mathnormal{A}=(2.97 \pm 0.114) \times 10^6~ (m.w.e)~ m^{-2} s^{-1}$. 

\begin{figure}[H]
	\centering
	\includegraphics[width=0.7\linewidth]{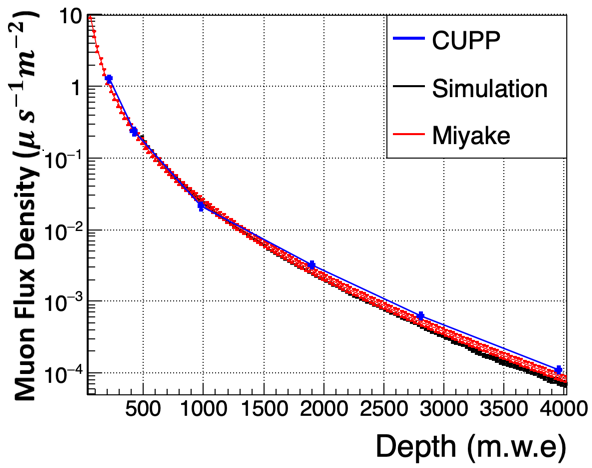}
	\caption{Muon flux density as a function of depth comparison of CUPP data points(blue), Geant4(black), and  Miyake fitted CUPP data(red).
 }
	\label{fig:muon_density}
\end{figure}

The muon cosmic ray flux density generated by the Geant4 Model and predicted by Miyake's formula are compared as a function of depth in Figure \ref{fig:muon_density}. The average difference is $5.8\%$ over the range, 500 m.w.e. to 1200 m.w.e.. The muon flux normalization 4 m rock above the experimental cavities are compared in Table \ref{ta:density_depth}. 
\begin{table}
\centering
\caption{Comparison of the Geant4 and Miiyake single parameter fitted Underground muon density, 4m rock above the experimental caverns.}

\begin{tabular}{|l|c|c|}
\hline

\begin{tabular}[c]{@{}l@{}}Depth \\ (m.w.e.)\end{tabular} & \begin{tabular}[c]{@{}c@{}}Miyake Fitting\\ muon s$^{-1}$ m$^{-2}$\end{tabular} & \begin{tabular}[c]{@{}c@{}}Geant4 Simulation\\ muon s$^{-1}$ m$^{-2}$\end{tabular} \\ \hline
571.6                                                     & 0.114                                                             & 0.123                                                                    \\ \hline
1154.6                                                    &  0.0154                                                         &  0.0148                                                                \\ \hline
\end{tabular}
\label{ta:density_depth}
\end{table}

\begin{figure}[!ht]
	\centering
	\includegraphics[width=1.0\linewidth]{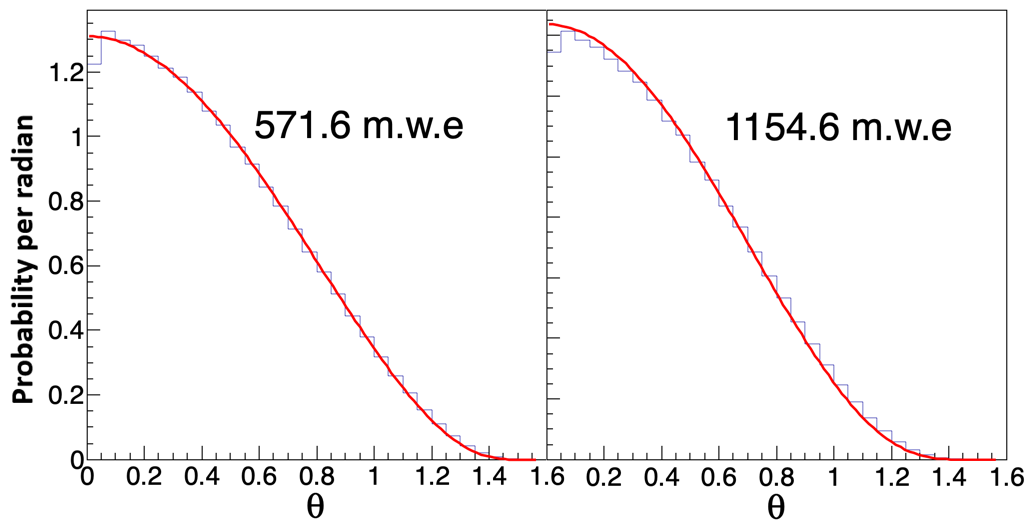}
        \caption{Comparison of the normalized cosmic ray muon angular distributions  4 m rock above the two experimental  halls.  Geant4 (blue).  Miyake's formula (red). }
	\label{fig:Angular}
\end{figure}

The muon angular distribution formula of Miyake \cite{kf2001cosmic} is given by, 
\begin{equation}
  I(X,\theta) = I(X,0^{\circ })cos^{1.53}(\theta)e^{-8.0\times10^{-4} X(sec(\theta)-1)}
  \label{equ:angle_dis}
\end{equation}
\noindent where $I(X,0^{\circ })$ is the vertical intensity at depth X. The underground muon angular distributions predicted by the Monte Carlo and Miyake's formula are compared at the two experimental depths in Figure \ref{fig:Angular}. At the two experimental locations the averaged rms difference between the two distributions is less than 1\%.

\subsection{Charge Ratio of the Surface Muons}
The final input to the Monte Carlo simulation is the charged muon flux ratio. At depth, for low energy muons, E$_{\mu}<$1 GeV, the  $\mu^-$ cross section with lead is larger than the $\mu^+$ due to attraction with the nucleus.  For higher energy muons, E$_{\mu}>$1 GeV, the cross sections is nearly charge independent.   This effect is displayed in Table \ref{ta:pm_compare}, which shows the average generated neutron number and the reaction probability for muons incident on a 30 cm cube lead target. The reaction probability is the probability of  generating at least one neutron.

To correctly take into account the effects of the muon flux charge ratio, the $\mu^+$ and $\mu^-$ Geant4 simulation are separately modeled. Only at the end of the complete simulation are they recombined using a single charge ratio value.  At 583 m.w.e. depth,  muons $<1$ GeV require a minimum energy of $\sim 150$ GeV at sea level to propagate to this level. Likewise at 1166 m.w.e. depth, an energy of $\sim 340$ GeV is required. Because the $\mu^+$ to $\mu^-$ ratio at the surface between 100 GeV and 500 GeV is constant, the value $1.29 \pm 0.13$\cite{khachatryan2010measurement} is selected. The estimated charge ratio for higher energy muons is less than the error of the selected value.

\begin{table}
\centering
\caption{The average number of produced neutrons excluding zero neutron production and the probability of generating at least one neutron for a $\mu^+$ or a $\mu^-$ incident on a 30 cm cube lead target. From the Geant4 simulation. }
\begin{tabular}{|l|l|c|c|c|}
\hline
 &  & 100 MeV   & 1 GeV    & 10 GeV    \\ \hline
\multirow{2}{*}{$\mu^-$} &  $<Neutron>$ & 4.9   & 4.7   & 6.2 \\
\cline{2-5} 
    & \begin{tabular}[c]{@{}l@{}}Reaction \\ Probability\end{tabular}         & 0.87     & 0.0013  & 0.012  \\ \hline
\multirow{2}{*}{$\mu^+$} & $<Neutron>$ & 1.4      & 4.3     & 6.5     \\ \cline{2-5} 
    & \begin{tabular}[c]{@{}l@{}}Reaction \\ Probability\end{tabular}  & 0.0034  & 0.0014  & 0.011  \\ \hline
\end{tabular}
\label{ta:pm_compare}
\end{table}

\section{Neutron Production Simulation}

\subsection{Geant4 Model}

All muon and secondary particle processes in the rock and lead are simulated with Geant4-11.01 using the physics list QGSP-BERT-HP.  This list was selected based on its use for LHC experiments. In particular ATLAS and CM have studied the physics performances of the physics lists and converged on the use of the QGSP-BERT physics list as the most comprehensive and thus the default \cite{kiryunin2006geant4}. Additional simulations for this study included FTFP-BERT-HP, QGSP-BIC-HP and QBBC for possible use in systematic error estimation. However the difference among FTFP-BERT-HP and QGSP-BIC-HP remained smaller than the statistical uncertainties, while there is a 15\% deviation in the QBBC list neutron production compared with the other physics lists, making the systematic error estimate using the variation in the physics lists not useful.

\subsection{Model Simulation Universe}

The Geant4 simulation begins full shower modeling starting in a horizontal layer 4m above the laboratory hall.  Muon generated showers are fully developed allowing the shower particles to interact with the lead target and detector system to generate neutrons. 

\begin{figure}[!ht]
\centering

\centering
\includegraphics[width=1.0\linewidth]{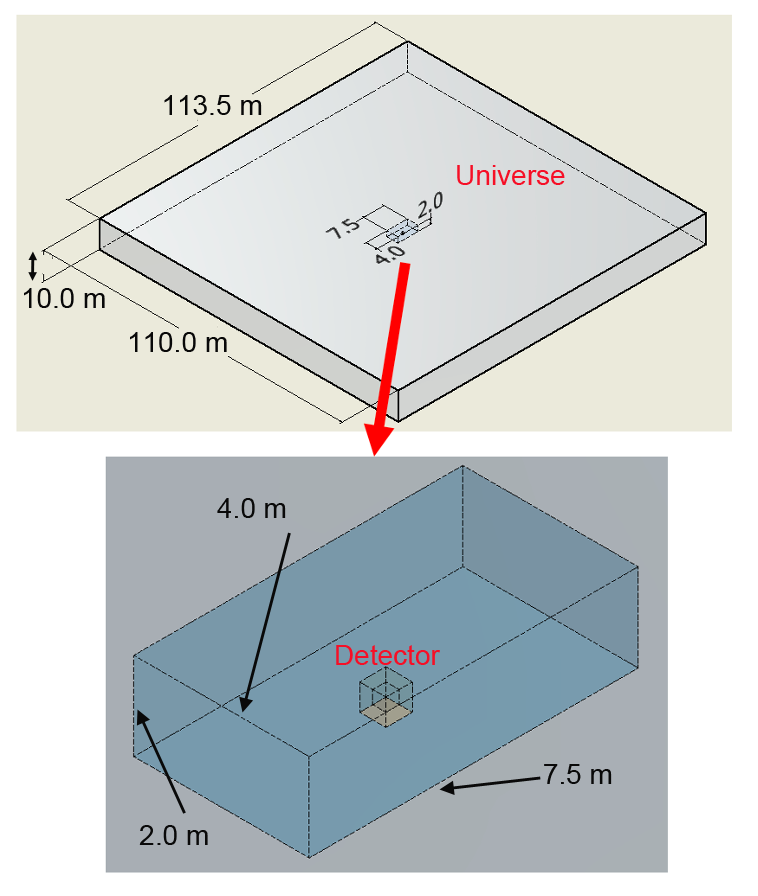}

\caption{(Upper) Proportional illustration of the Geant4 Universe, includes the rock layer and (Lower) the laboratory cavern and the detector system, 30 cm Pb-cube inside 60 cm poly-cube sitting centered on the cavern floor.}
\label{fig:Model_geo}
\end{figure}

The model universe is displayed in Figure \ref{fig:Model_geo}. The NMDS-II detector system, is centered on the floor inside a cavity ($7.5 m \times 4 m \times 2 m$) surrounded by rock. The horizontal thickness of each of the universe's rock side walls is 53 meters.  The cavity rests on and is capped by a 4m thick rock floor and  roof. Both extend to the outer edges of the side walls forming a rectangular cube. The dimensions where chosen to assure cosmic ray muons with angles from 0 degree to 85.5 degree intersect the top of the universe's roof. The number of slant angle muons passing through the target not covered by this angular range is smaller than 1 over the course of the experiment. 
The Geant4 simulation inputs are then, (1) the muon flux intensity at depth, (2) the correlated muon energy and muon angular distribution at depth and finally, (3) the cosmic ray muon charge ratio.
The full simulation starting point is the universe's top surface or roof, $113.5 m \times 110 m$. The total number of muon events intersecting the universe's roof in the experiments live times are shown in Table \ref{ta:Normalization}. The values are found using Eq.~\ref{equ:depth}, Miyake's fit to the experimentally measured \cite{enqvist2005measurements} muon flux density values, shown in Table \ref{ta:density_depth}.   
In order that the simulations have superior statistics compared to the data, more than 10 times the experimental statistics were simulated, also displayed in Table \ref{ta:Normalization}.

\begin{table}
\centering
\caption{Number of cosmic ray muons intersecting the roof of the Geant4 universe during the experimental live times and the number of simulated events at each detector depth. \\ }

\begin{tabular}{|l|c|c|}
\hline
             & 583 m.w.e. & 1166 m.w.e. \\ \hline
Live Time hrs & 6504 & 1440\\ \hline
Experiment   & $3.33\times 10^{10}$ & $9.97 \times 10^8$  \\ \hline
Simulation   & $4.58\times 10^{11}$ & $2.29\times 10^{11}$  \\ \hline
\end{tabular}
\label{ta:Normalization}
\end{table}

\begin{figure}
\centering
\includegraphics[width=0.8\linewidth]{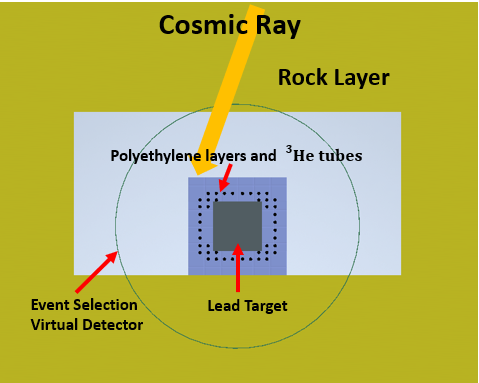}
\caption{Illustration of the Geant4 Event Selection Virtual Detector, relative to the $^3He$ detector system and Pb-target centered on the cavern floor. The dimensions are not in proportion.}
\label{fig:Model_illustration}
\end{figure}

\subsection{Full Event Simulation Trigger}

Because of the large size of the simulation universe and the corresponding low interaction rate of cosmic ray showers with the experiment's detectors, a simulation trigger was developed to further reduce the required computational run time. Once the simulation located a muon on the roof of the universe, using its angular parameters the muon track was pointed at the target. The intersection of the muon track with a spherical surface virtual detector was checked and the event was rejected from further consideration if not intersecting the virtual detector. This virtual detector forms a MC trigger for which muons will be fully simulated as they pass through or nearby the target. Figure \ref{fig:Model_illustration} illustrates the geometry of the model.

For an accurate simulation, it is critical that the event selection virtual detector should be given an appropriate size. Otherwise,  some events with neutrons detected will be mis-rejected if the size is too small. However, computing time is roughly proportional to the volume of the virtual detector, or $R^3$. Too much computing time will be taken if the event selection virtual detector is too large.

\begin{figure}
\centering

\begin{subfigure}
\centering
\includegraphics[width=1.0\linewidth]{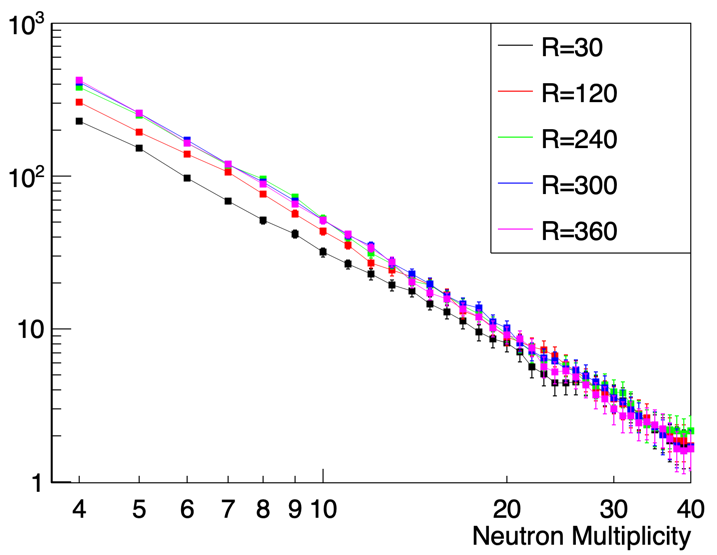}
\end{subfigure}

\caption{Log events detected as a function of Log detected neutron multiplicity (n) for selected virtual detector radii.  R=30 cm separates from others curve at n = 24. R = 120 cm separates from others at n=13, and R=240 cm separates at n=4. The curve R=300 and R=360 cm are still merged together at n=4. }
\label{fig:VD_size}
\end{figure}

To optimize the radius of the virtual trigger detectors, simulations of the number of detected neutrons as a function of the virtual detector's radius were studied. Figure \ref{fig:VD_size} displays the results. At multiplicity larger than 25 observed neutrons ($\sim 120$ produced neutrons), all the curves for radii between 30 cm to 360 cm are not distinguishable.  For these events the distance between the center of the target and the extension line of the muon's initial direction is smaller than 30 cm. When the observed neutron multiplicity is from 14 to 25, the  R=30 cm curve is separated from other curves. In the same Figure \ref{fig:VD_size}, the R=120 cm curve  is separated from other curves at observed multiplicity equals to 13. At neutron multiplicity equals to 4, the R=240 cm curve  is separated from the curves R=300 cm and R=360 cm, while the  R=300 cm and R=360 cm  curves are still merged. 

Given these distributions, the virtual event trigger detector radius is selected to be R=300 cm as neutron multiplicity events lower than 4, are not consider in the simulation nor in the experimental measurement.

\subsection{Simulation Statistical Errors Using Bootstrapping}

Finally, once a muon is accepted by the virtual trigger for full shower simulation and that event's shower simulation is completed, its neutrons hit times on the $^3He$ tubes were re-sampled 100 times to get the average response of the NMDS-II He-3 neutron detector system to that event. The statistical errors for the simulations detected neutron multiplicity are estimated using a Bootstrap method\cite{johnson2001introduction}.
\section{Experimatal and Simualtion Data Analysis}

\subsection{Power Law $\chi^2$ Fits}

\begin{figure}
\centering

\centering
\includegraphics[width=1.0\linewidth]{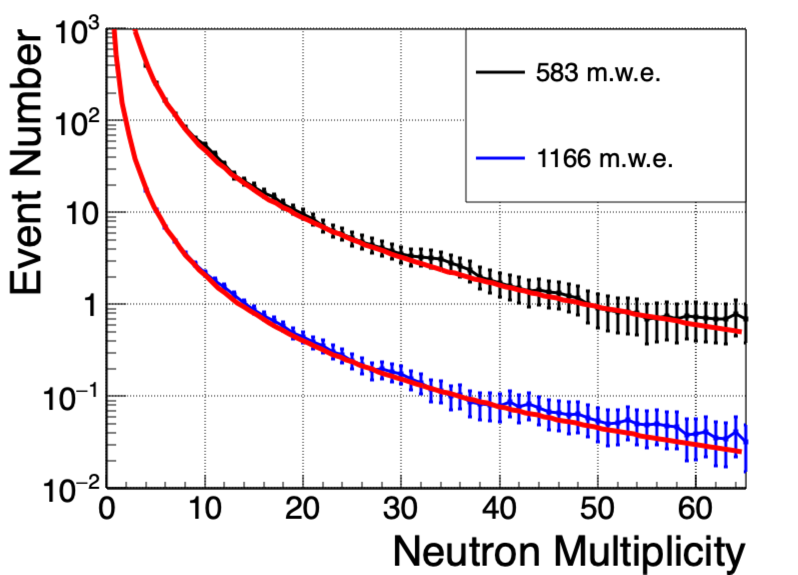}

\caption{  Comparison of the simulated observed neutron multiplicity distributions with a two parameter power law fit at 583 m.w.e.  and at 1166 m.w.e..}
\label{fig:simu_fit}
\end{figure}

The MNDS-II Geant4 simulations results are displayed in Figure \ref{fig:simu_fit}, for events with observed multiplicity of 4 or more neutrons. The neutron multiplicity distributions are expected to have a power-law distribution.  This occurs because the initiating primary hadronic space component, having energy above a few GeV, has a power law intensity fall off, E$^{-\gamma}$ with $\gamma\sim2.7$\cite{apanasenko1999primary}. Because of this, the secondary high energy pion and produced daughter muons have energy intensity fall offs with a similar $\gamma$ value \cite{grieder2001cosmic2}. As shown in Eq. \ref{equ:Surfcosmic}, again the underground muon  energy intensity falls off with a similar $\gamma$ value.  From this the muon induced energy deposited in the Pb-target should have a power law fall off.  And finally, because the average number of neutrons produced is roughly proportional to the energy deposited, again the neutron multiplicity is expected to have a power law fall off.   

This reasoning is confirmed in Figure \ref{fig:simu_fit} displaying the two parameter power law fits, $y = k\times n^{-p}$ to the simulations of the neutron multiplicity distributions.   Where k is the amplitude parameter, n is the neutron multiplicity and p is the power law exponent of n. 

The results are $\chi^2$/DoF = 1.24 for the 583 m.w.e. simulation  and $\chi^2$/DoF = 1.24 for the 1166 m.w.e. simulation. Both fits have 60 degrees of freedom.  The fitting parameters are displayed in Table \ref{ta:fit_para}. 
The same power law parameter value is found in solar neutron emissivity \cite{powerlaw_Solar}.

\begin{figure}
	\centering
	\includegraphics[width=1.0\linewidth]{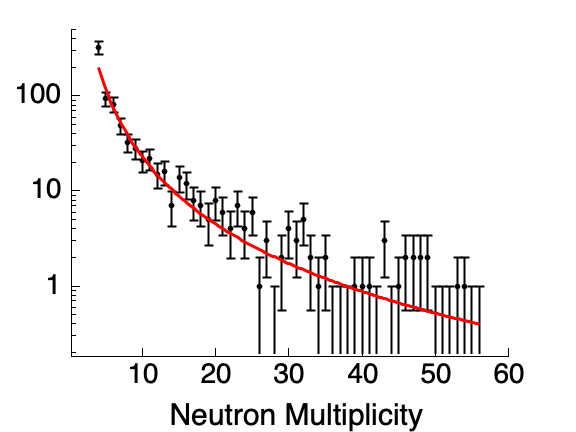}
	\caption{Experimental data neutron multiplicity distribution  at 583 m.w.e. fit with a 2 parameter  power law function. Bins with zero counts were fit with an error of 1. }
	\label{fig:583_exp_fit}
\end{figure}

The experimental data at  583 m.w.e. is compared to its power law fit in the Figure \ref{fig:583_exp_fit} and its parameters are displayed in Table \ref{ta:fit_para}. In fitting, a multiplicative 5.6\% systematic error at each  point was included and  placed in quadrature with the statistical error.  The additional error is due to deterioration of the $^3$He tube over the course of data collection. 
As was the case with the simulation, the data also fits well to a power law yielding $\chi^2$/DoF = 0.76 for 55 degrees of freedom.  

\subsection{ Power Law Maximum Likelihood Fit}

For the data at 1166 m.w.e. a Maximum Likelihood Estimation (MLE) method is used to measure the power law parameters because there are only 6 points with multiplicity above 5 observed neutrons in the data set. The measurement is made for events whose neutron multiplicity is  equal to or larger than 4. In this case the multiplicity probability distribution is $D = A\times n^{-p}$ ($n \geqslant 4$) normalized to one, making the k-parameter or amplitude parameter p-dependent.  The one parameter fit then yields p=2.50 with an only slightly asymmetric Gaussian shaped Likelihood distribution having  $\sigma_{RMS} = $  0.35. From this p = $2.50 \pm 0.35$. The variation in k is then calculated from the variation in p  and is displayed in Table \ref{ta:fit_para}.  

All the fitting results are summarized in Table \ref{ta:fit_para}.

\begin{table}
\centering
\caption{ Geant4 simulation and  NMDS-II data, neutron multiplicity distribution  power law function, $k\times n^{-p}$, fit parameters. The fits are over the range [4,60] observed neutrons. All p parameters can be compared. k values are dependent on the data collection time and are to be compared only at the same depth. } 

\begin{tabular}{|l|c|l|c|c|c|}
\hline
\begin{tabular}[c]{@{}l@{}}Depth\\ (m.w.e.)\end{tabular} & \begin{tabular}[c]{@{}c@{}}Time\\ (hour)\end{tabular} &            & p                                                           & \begin{tabular}[c]{@{}c@{}}k\\ ($10^3$)\end{tabular}    & \begin{tabular}[c]{@{}c@{}}$\chi^2$ \\per DoF\end{tabular} \\ \hline
\multirow{2}{*}{583}       & \multirow{2}{*}{6504}       & Geant & \begin{tabular}[c]{@{}c@{}}2.37\\ $\pm$ 0.01\end{tabular}   & \begin{tabular}[c]{@{}c@{}}11.6\\ $\pm$ 2.8 \end{tabular}    & 1.24             \\ \cline{3-6} &              & Exper & \begin{tabular}[c]{@{}c@{}}2.36 \\ $\pm$ 0.10\end{tabular}  & \begin{tabular}[c]{@{}c@{}}5.2\\ $\pm$ 1.2\end{tabular}     & 0.76   \\ \hline
\multirow{2}{*}{1166}      & \multirow{2}{*}{1440}       & Geant & \begin{tabular}[c]{@{}c@{}}2.31 \\ $\pm$ 0.01\end{tabular} & \begin{tabular}[c]{@{}c@{}}0.44\\ $\pm$ 0.11 \end{tabular}   & 1.24      
\\ \cline{3-6} 
&    & Exper & \begin{tabular}[c]{@{}c@{}}2.50\\ $\pm$ 0.35\end{tabular}   & \begin{tabular}[c]{@{}c@{}}0.19\\ $+0.17 / -0.10$\end{tabular} & 0.76                                                       \\ \hline
\end{tabular}
\label{ta:fit_para}
\end{table}

\section{Data and Simulation Comparison}
\subsection{Statistical Error Comparison}
\begin{figure}
\centering

\begin{subfigure}
\centering
\includegraphics[width=0.9\linewidth]{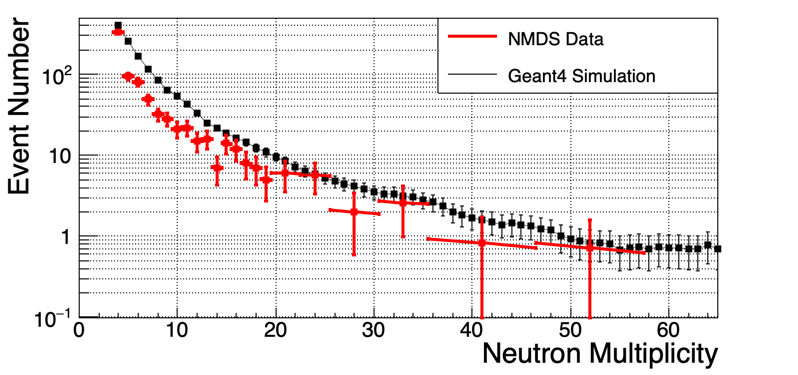}
\end{subfigure}

\begin{subfigure}
\centering
\includegraphics[width=0.9\linewidth]{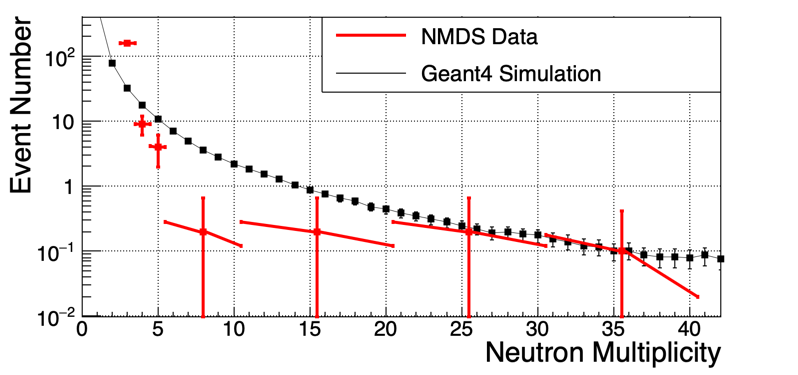}
\end{subfigure}
\caption{Comparison of the NMDS-II (Red)  neutron multiplicity distributions to the normalized Geant4 simulation (black);  (Upper) normalized to 6504 hours of data collection at 583 m.w.e. and (Lower) 1440 hours of data collection at 1166 m.w.e.}
\label{fig:compare_rebin}
\end{figure}

The comparison between the NMDS-II data and Geant4 simulation for 583 m.w.e. and 1166 m.w.e. are shown in Figure \ref{fig:compare_rebin}.
Comparing the fit parameter results in Table \ref{ta:fit_para} shows that the power law parameter p values at both depths and for the data and simulation are in excellent agreement.  However, the k or amplitude  parameter, are in disagreement.  The k parameter can only be compared between simulation and data at the same depth. k depends on the number of hours of data collection and the muon flux.  The Geant4 model yields $\sim$2 times as many events as the data at 583 m.w.e.,  and $\sim$4 times the number of events at 1166 m.w.e.. Because a power law function is scale invariant in the power p,  this difference  can be accounted for by a scale shift in the simulated mean neutron multiplicity of 35\% at 583 m.w.e. and 82\% at 1166 m.w.e..

These same disagreements are found in data collected at the Boulby Underground Laboratory in UK \cite{araujo2008measurements} and Tubingen Shallow Underground Laboratory in Germany \cite{KNEIL201987}. In the Boulby experiment, there was a 0.82 $m^3$ liquid scintillator at a depth of 1070 m, or 2850 m.w.e.. Their Geant4 model(8.2) gave an average neutron per muon 1.8 times higher than the observed value \cite{araujo2008measurements}, yielding a $k$ value $\sim 4.2$ times higher than the observed. 

To make a full comparison the systematic errors in the experimental data and simulation must be taken into account.

\subsection{Experimental Systematic Errors}

The experimental systematic errors are displayed in Table \ref{ta:systematic_error_summary}. In order, the Pyhas\"almi(CUPP) mine measurement for the cosmic ray flux normalization is taken from Reference \cite{enqvist2005measurements}, stated to be 10\%. The error in the rock over burden thickness is estimated to be $\sim 20\  m.w.e$.  This over burden error is converted to a change in the cosmic flux calculated using the Miyake formula  Eq. \ref{equ:depth} to be $9.0\%$.

The efficiency of the $^3He$ neutron detection system is $(23.2\pm 1.4)\%$ estimated by the V. G. Khlopin Radium Institute's group \cite{CUPP2005}. This error causes a scale shift in the simulated observed neutron number. The fit p-parameter in the simulation is used to estimate the shift.
In addition, there is a 1.2\% efficiency decrease for the $^3He$ tubes caused by the degradation of the tubes over the course of the measurements. This systematic error causes a scaling effect in the data and was included in the data fitting by increasing each data points error, as was discussed, so is not included again here. Finally, the error in the experimental live time is negligible compared with cosmic ray flux error and depth error, so is dropped. 

These experimental errors, added in quadrature, result in a total systematic error of only $\sim$20\% which does not significantly improve the comparison between the data and the simulations. Thus, the experimental systematic erros can not account for the difference between the data and the simulation. 

The summary of systematic errors in the NMDS-II experiment is shown in the Table \ref{ta:systematic_error_summary}.

\begin{figure}
\centering

\begin{subfigure}
\centering
\includegraphics[width=1.0\linewidth]{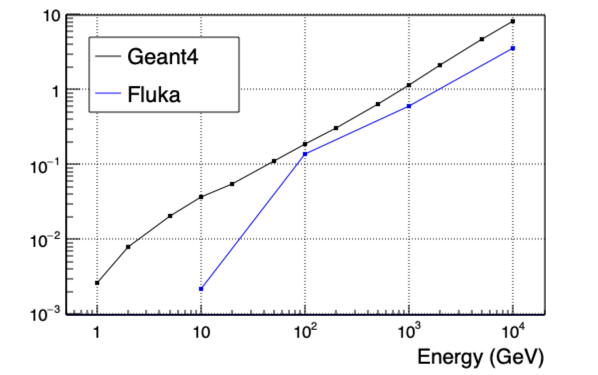}
\end{subfigure}

\begin{subfigure}
\centering
\includegraphics[width=1.0\linewidth]{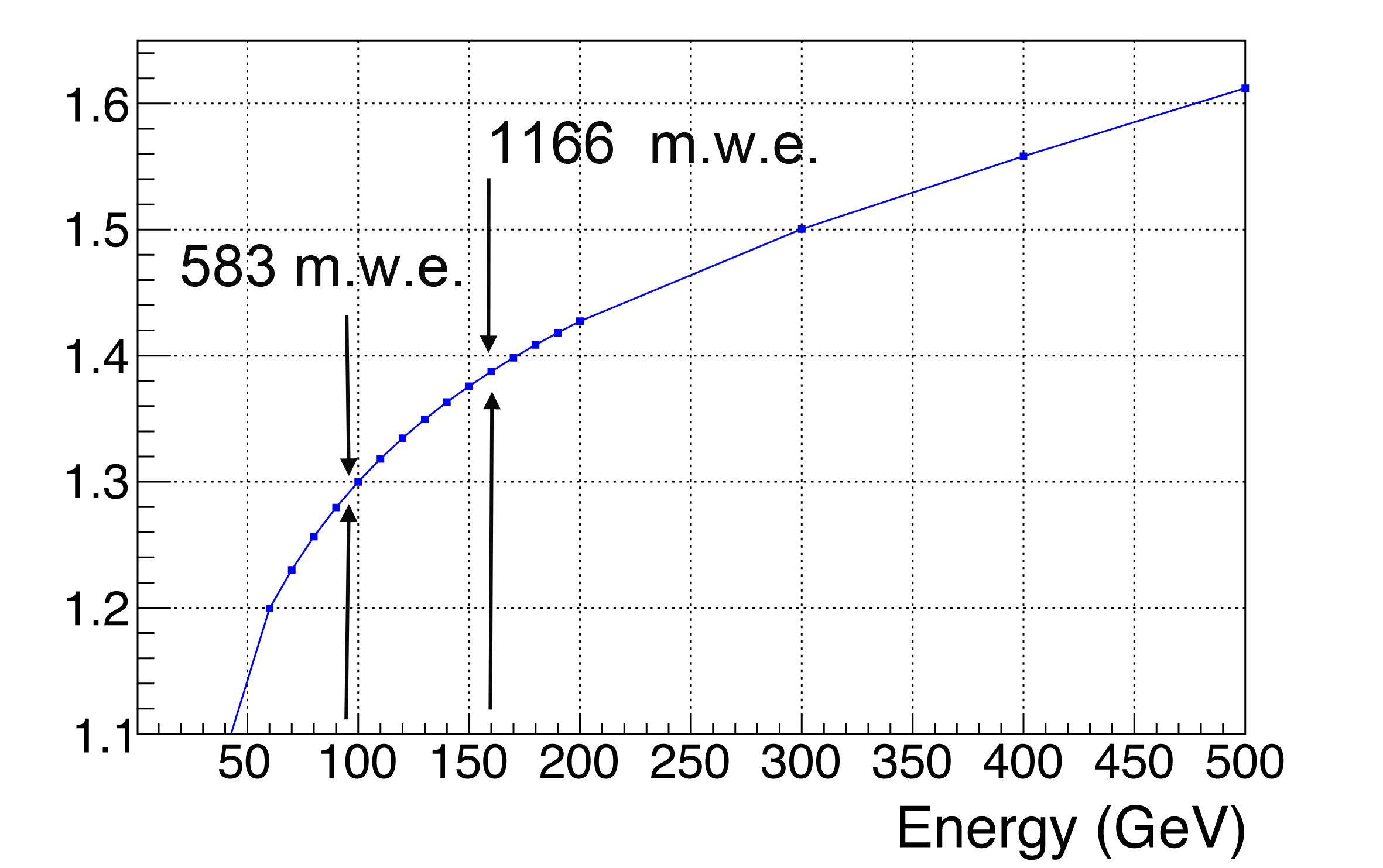}
\end{subfigure}

\caption{(Upper) Average number of neutrons produced as a function of incident muon energy from Geant4 and FLUKA.  (Lower) Average neutron number ratio, Geant4/FLUKA as a function of muon energy. Muon mean energy at 583 m.w.e. is $\sim$ 95 GeV, with corresponding ratio 1.29. Muon mean energy at 1166 m.w.e. is $\sim$ 160 GeV, with corresponding ratio 1.39.}
\label{fig:Compare_Simus}
\end{figure}
\subsection{Error in the Geant4 Model}

To estimate the errors in the Geant4 neutron production modeling, a separate simulation of muon induced neutron production is performed comparing Geant4 and FLUKA. In the study, a muon with fixed initial energy is incident normal to a 30 cm-thick, infinity large lead sheet. The number of neutrons produced are then recorded and averaged over the event sample.  The Geant4 and FLUKA simulated neutron production as a function of muon energy is shown in Figure \ref{fig:Compare_Simus} (Upper). In Geant4, muons produce more neutrons than in FLUKA. The production difference increases as a function of muon energy which is clearly seen in the ratio R= Geant4/FLUKA displayed as a function of incident muon energy in Figure \ref{fig:Compare_Simus} (Lower).  If the difference between the two Monte Carlo simulations is taken as the systematic error in the Geant4 simulation, then this systematic error in produced neutron number results in a scaling error. Due to the scaling invariance of the power law function index, the scaling of k is calculated using the average muon energy incident on the Pb-target at each depth. From the Geant4 cosmic ray muon simulations the muon mean energy at 583 m.w.e. is $\sim$ 95 GeV, with corresponding ratio,  R = 1.29. The muon mean energy at 1166 m.w.e. is $\sim$ 160 GeV, with corresponding ratio R = 1.39.  The neutron number scaling factor S is then calculated,
\begin{equation}
  {S=1-R^{-p},}  
\label{eq:scaling}
\end{equation}
and the resulting errors are displayed in Table \ref{ta:systematic_error_summary}.

\subsection{Comparison Including all Errors}
\begin{figure}
	\centering
	\includegraphics[width=0.9\linewidth]{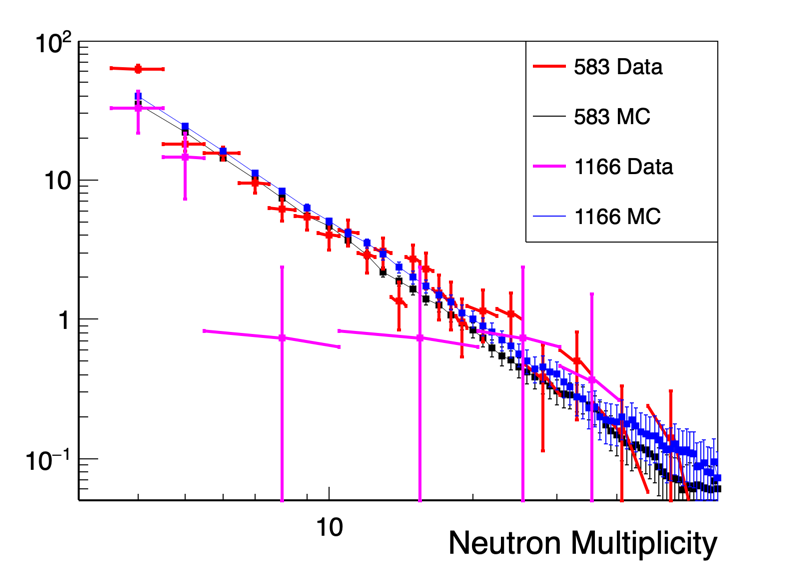}
	\caption{Amplitude independent comparison of the simulations and experimental neutron multiplicity distributions at both depths. Each curve is fit using $k\times n^{-p}$, however k is shifted to 1000, an arbitrary number, for all distributions for comparison of the power law slope parameter p-values.}
	\label{fig:NoK}
\end{figure}

In Table \ref{ta:fit_para} and displayed in Figure \ref{fig:NoK} the neutron multiplicity power law parameter p for both the simulation and the data are in good agreement and give good fits to a power law as is expected due to the power law spectrum of the initiating  hadronic space component. Errors due to cosmic ray flux intensity, lab depth and $^{3}$He tube efficiency, shown in Table \ref{ta:systematic_error_summary}  do not effect the power law index due to its scaling invariance.  These errors cause only an overall normalization motion or amplitude motion in the  k-parameter.  Likewise, in the  simulation, these same errors  produce an error only in the k value.  In addition any error in Geant4 neutron generation, again displayed in Table \ref{ta:systematic_error_summary}, produces no change in the power law index, the p-parameter, due to its scale invariance.  From this, taking the p-values shown in Table \ref{ta:fit_para} and forming a ratio between the data and the simulation at a given depth, does not require the inclusion of the systematic errors as they divide out. The resulting comparison at 583 m.w.e. is 1.00~$\pm$ 0.04 and at 1166 m.w.e. is 1.08 $\pm$ 0.15.

As discussed, comparing the amplitude parameters, shown in Table \ref{ta:fit_para} of the simulation and data yeilds a disagreement.  The experimental systematic errors can not account for this disagreement.  However, assuming the difference between the Geant4 and FLUKA simulations are a measure of their systematic error, Table \ref{ta:ratio_results} shows that the ratio Ratio = Greant4/data -1 is within 1-standard deviation agreement.

The large size of the systematic error in the simulated neutron production, also shown in Figure \ref{fig:Compare_Simus}, indicates that the simulations are poor indicators of the amplitude of muon induced neutron production, making the measured NMS-II neutron multiplicity data critical input to normalizing cosmic ray induced neutron production at depth.

\begin{table}
\caption{Systematic errors in the data and Geant4 simulations.  The detector degradation is included in the data fit. }
\label{ta:systematic_error_summary}
\begin{tabular}{|lcc|}
\hline
\multicolumn{3}{|c|}{\begin{tabular}[c]{@{}c@{}}Errors Effecting p for the \\ NMDS-II Experimental Data\end{tabular}}                                                                                                          \\ \hline
\multicolumn{1}{|c|}{Error Source}                                                             & \multicolumn{1}{c|}{Motion Direction} & \begin{tabular}[c]{@{}c@{}}Systematic Error\\ Propagated to \\ Event Number\end{tabular} \\ \hline
\multicolumn{1}{|c|}{\begin{tabular}[c]{@{}c@{}}1.2 \%\\ He-3 Tube\\ Degradation\end{tabular}} & \multicolumn{1}{c|}{{\begin{tabular}[c]{@{}c@{}}neutron\\ number\end{tabular}}}                & \begin{tabular}[c]{@{}c@{}}583 : 13.0\%\\ 1166 : 12.7\%\end{tabular}                     \\ \hline
\multicolumn{3}{|c|}{\begin{tabular}[c]{@{}c@{}}Errors Effecting k for the \\  Simulation or data\end{tabular}}                                                                                                                  \\ \hline
\multicolumn{1}{|l|}{\begin{tabular}[c]{@{}l@{}}10\% of\\ Cosmic Ray Flux\end{tabular}}        & \multicolumn{1}{c|}{{\begin{tabular}[c]{@{}c@{}}event\\ number\end{tabular}}}              & \begin{tabular}[c]{@{}c@{}}583: 10\%\\ 1166: 10\%\end{tabular}                           \\ \hline
\multicolumn{1}{|l|}{\begin{tabular}[c]{@{}l@{}}20 m.w.e. of\\ Lab Depth\end{tabular}}         &\multicolumn{1}{c|}{{\begin{tabular}[c]{@{}c@{}}event\\ number\end{tabular}}}                & \begin{tabular}[c]{@{}c@{}}583 : 8.3\%\\ 1166: 5.3\%\end{tabular}                        \\ \hline
\multicolumn{1}{|l|}{\begin{tabular}[c]{@{}l@{}}1.4\% He-3\\ Counter Efficiency\end{tabular}}  & \multicolumn{1}{c|}{{\begin{tabular}[c]{@{}c@{}}neutron\\ number\end{tabular}}}               & \begin{tabular}[c]{@{}c@{}}583: 13.8\%\\ 1166 : 13.4\%\end{tabular}                      \\ \hline
\multicolumn{1}{|l|}{Geant4 Model}                                                             & \multicolumn{1}{c|}{{\begin{tabular}[c]{@{}c@{}}neutron\\ number\end{tabular}}}                & \begin{tabular}[c]{@{}c@{}}583: 45\%\\ 1166: 53\%\end{tabular}                           \\ \hline
\multicolumn{1}{|l|}{\textbf{Total Systematic}}                                                                    & \multicolumn{1}{l|}{}                 & \begin{tabular}[c]{@{}c@{}}\textbf{583: 49\%}\\ \textbf{1166: 56\%}\end{tabular}                           \\ \hline
\end{tabular}
\end{table}

\begin{table}
\caption{Comparison between Geant4  and collected data values for the amplitude k-parameters.  For Geant4, statistical and systematic are displayed separately. For the data sets all systematic and statistical errors have been added in quadrature.  Ratio = Greant4/data  with all errors added in quadrature.}
\label{ta:ratio_results}
\centering
\begin{tabular}{|l|c|c|c|}
\hline
\begin{tabular}[c]{@{}l@{}}Depth\\ (m.w.e.)\end{tabular} & \begin{tabular}[c]{@{}c@{}}Geant4\\ k ($10^3$)\end{tabular}       & \begin{tabular}[c]{@{}c@{}}NMDS-II \\ k ($10^3$)\end{tabular}   & $Ratio-1$                                                    \\ \hline
583                                                      & \begin{tabular}[c]{@{}c@{}}11.6 \\ $\pm 2.8  \pm 5.7$ \end{tabular}  & \begin{tabular}[c]{@{}c@{}}5.2 \\ $\pm 1.2$ \end{tabular}         & \begin{tabular}[c]{@{}l@{}}1.2\\ $\pm 1.3$ \end{tabular}     \\ \hline
1166                                                     & \begin{tabular}[c]{@{}c@{}}0.44 \\ $\pm 0.11 \pm 0.25$ \end{tabular} & \begin{tabular}[c]{@{}c@{}}0.19 \\ + 0.17/ -0.10\end{tabular} & \begin{tabular}[c]{@{}c@{}}1.3\\ +2.6/ -1.9\end{tabular} \\ \hline
\end{tabular}
\end{table}

\section{Conclusion}

The GEANT4 prediction and the observed data neutron event multiplicity distributions have matching power law shapes $k\times n^{-p}$.  The neutron multiplicity production amplitude for the experimental data and simulation are not in agreement and can not be explained by the systematic error in the experimental measurement. Because of this disagreement the simulation does not allow direct prediction of the expected event counts. However, the power law exponent or  index values from both simulations and from both experimental data sets are in excellent agreement.

\section*{Acknowledgements}
 The authors thank Dr. Thomas Ward from TechSource, Inc for providing the NMDS-II experimental data sets and many insightful discussions, and Alex Barzilov, University of Nevada, Las Vegas, for discussions concerning the $^3$He-detectors. 

 This work was funded in part by a grant from the U.S. Department of Energy Office of Nuclear Energy, Contract No. DE-SC0007884; In part by The Department of Physics and Astronomy at Purdue University; and in part by TechSource Inc.

\appendix

\section{Muon Energy Loss in the Rock} \label{app:muon_propogation}

 Muons lose energy by ionization and by three radioactive processes; bremsstrahlung, production of electron-position pairs and photonuclear interactions. The muon energy loss was calculated using the form \cite{CosmicRayPDG}, 
 
\begin{equation}
-\frac{dE_\mu}{dX} = a + bE_{\mu},
\label{equ:ELoss}
\end{equation}

\noindent where $\mathnormal{a}$ is the ionization loss and $\mathnormal{b}$ is the sum of the three fractional radiative losses.  Both are slowly varying functions of muon energy as shown in Table \ref{ta:MuonELoss} for standard rock.
A second-order polynomial was used to fit $\mathnormal{a}$ and $\mathnormal{b}$, each independently, as a function of $log_{10}E(GeV)$, yielding the fits values,
\begin{equation}
a = - 0.005 * (log_{10}E)^2 + 0.277*log_{10}E + 1.9
\label{equ:ELoss_a}
\end{equation}
\begin{equation}
b = - 0.1775 * (log_{10}E)^2 + 1.7105*log_{10}E + 0.3575    
\label{equ:ELoss_b}
\end{equation}
with only 1 degree of freedom.  These values are input in Equation \eqref{equ:ELoss} yielding the point-by-point transformation of the surface energy distribution to one at depth.

\begin{table}[H]       
    \centering
    \caption{Average muon range R and energy loss parameters calculated for standard rock\cite{CosmicRayPDG}. \\} 

    \scalebox{0.87}
    { 
    \begin{tabular}{|c|c|c|c|c|c|c|c|c|}
    \hline
     $\mathbf E_{\mu}$ & \bf R & \bf a & $\mathbf{b_{brems}}$ & $\mathbf{b_{pair}}$ & $\mathbf{b_{nucl}}$  & $\mathbf{bE_\mu / a}$ & $\mathbf{\delta E\ water}$\\
    \hline
    \bf GeV & \bf km.w.e & $\mathbf{MeV} $ &\multicolumn{3}{|c|}{$\mathbf{10^{-6}}$} & &$\mathbf{GeV} $   \\
    &&$\mathbf{g^{-1} cm^2} $& \multicolumn{3}{|c|}{$\mathbf {g^{-1} cm^2}$} &&$\mathbf{m^{-1}} $\\
    \hline
    10    & 0.05   & 2.17   & 0.70 & 0.70 & 0.50  & 0.0088 & 0.219\\
    100   & 0.41   & 2.44   & 1.10 & 1.53 & 0.41  & 0.1246 & 0.274\\
    1000  & 2.45   & 2.68   & 1.44 & 2.07 & 0.41  & 1.463  & 0.660\\
    10000 & 6.09   & 2.93   & 1.62 & 2.27 & 0.46  & 14.85  & 4.643\\
    \hline
    \end{tabular}
    }
  \label{ta:MuonELoss}
\end{table}

 The validity of this formulation when propagated to NMDS-II depth is supported by a comparison of the formulation and measured data \cite{CosmicRayPDG}, displayed together in Table \ref{ta:MuonR}.

\begin{table}[H]                            
   \centering
    \caption{Comprision between the measured range and predicted average muon range using Formula \eqref{equ:ELoss}, units in m.w.e.. \\} 

       \scalebox{0.9}
    { 
   \begin{tabular}{|c|c|c|c|}
   \hline
   $\mathbf E_{\mu}$ & \bf Measured & \bf Predicted& \bf Difference\\
    \hline
    10    & 50    & 48.7  & 2.6  \% \\
    100   & 410   & 409   & 0.24 \% \\
    1000  & 2450  & 2455  & 0.21 \% \\
    10000 & 6090  & 6500  & 6.3  \% \\
    \hline
    \end{tabular}
    }
  \label{ta:MuonR}
\end{table}

 The predicted average muon range in Table \ref{ta:MuonR} is obtained by numeral calculations. The muon path in the rock is divided into many small steps. Muon energy loss at each step is then
\begin{equation}
    \Delta E =\frac{dE}{dx}\cdot  \Delta x
\end{equation}

The length of each step was taken to be $\Delta x = 0.01$ m.w.e., and $\frac{dE}{dx}$ is found from Equation \eqref{equ:ELoss}. The energy of the muon is changed in each step, so that \textbf{a} in \eqref{equ:ELoss_a} and \textbf{b} in  \eqref{equ:ELoss_b} are also changed. The predicted muon average range in Table \ref{ta:MuonR} is the total length of the muon path in the rock when the energy becomes 0.

\begin{figure}[H]
	\centering
	\includegraphics[width=0.95\linewidth]{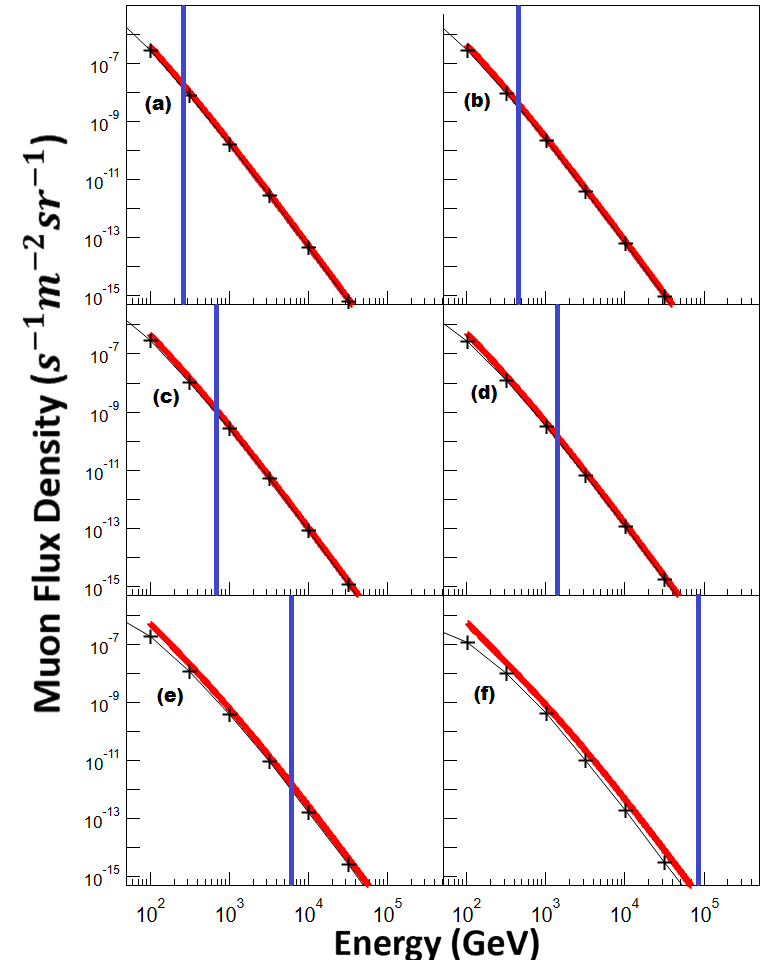}
	\caption{Sea level cosmic ray muon flux density at a fixed cos($\theta$) (a) 0.6, (b) 0.4, (c) 0.3, (d) 0.2, (e) 0.1, (f) 0.05, comparing Eq. \ref{equ:Surfcosmic} (red line) and  the measured data (black points). The comparison is to be made at energies above the the vertical bar (blue) indicating the muon energy cut-off or range at the experiments depth of 583 m.w.e..
}
	\label{fig:compare}
\end{figure}

The slant angle calculated flux and experimental flux is in agreement in the high energy region below the muon cutoff energy or range, shown as a horizontal line (blue) in Figure \ref{fig:compare} and listed in Table \ref{ta:MuonECut}. The highest energy events estimated to be observed by the detector is $\sim$ 10 TeV, limiting the slant angle to cos($\theta)> $0.05. The curvature of the Earth was taken into account in these calculations.

Figure \ref{fig:compare} shows the cosmic ray muon spectrum predicted by the sea level muon density Formula \eqref{equ:Surfcosmic} and from measured data  \cite{lipari1991propagation} as a function of cos$\theta$.  cos$\theta$ is chosen as 0.6, 0.4, 0.3, 0.2, 0.1 and 0.05. Only the muons in the part to the right of the blue bar are able to pass through the rock and propagate to 583 m.w.e.. Thus from the Figure \ref{fig:compare}, the muon flux density predicted by Formula \eqref{equ:Surfcosmic} matches well with the experimental data. 

 \begin{table}
  \begin{center}
    \caption{Muon energy cutoff at 583 m.w.e as a function of muon slant angle.  Eq. \ref{equ:Surfcosmic} is known to be valid for E$_{\mu} >$ 100 /cos($\theta$) GeV \cite{CosmicRayPDG}. The $E_{\mu}$ cutoff at each incident angle is the energy at which the muon exactly loses all the energy when passing through $583/ cos\theta$ m.w.e.. The curvture of the earth is considered in the calculation of the cutoff energy. \\} 
      \scalebox{0.9}
    { 
    \begin{tabular}{|c|c|c|c|}
    \hline
    $\mathbf{ cos ~ \theta}$ & $\mathbf{\theta}$ & $\mathbf{ 100 / cos \theta}$ &\bf{E$_\mu$ Cutoff }\\
    &&GeV&GeV\\
    \hline 
    1.00  & $0.0^\circ$  & 100  & 149\\
    0.60  & $53.1^\circ$ & 167  & 275\\
    0.40  & $66.4^\circ$ & 250  & 463\\
    0.30  & $72.5^\circ$ & 333  & 695\\
    0.20  & $78.5^\circ$ & 500  & 1330\\
    0.10  & $84.3^\circ$ & 1000 & 6045\\
    0.05  & $87.1^\circ$ & 2000 & 87800\\
    0.00  & $90.0^\circ$ &  -   & $> 10^6$\\
    \hline
    \end{tabular}
    }

  \label{ta:MuonECut}
  \end{center}
\end{table}

Although Formula \eqref{equ:Surfcosmic} is only valid when $E_\mu > 100/cos\theta\ GeV$ $\theta < 70^\circ$, agian, muons not in this region will not pass through the thick rock layer to a depth of 583 m.w.e.. From this the underground cosmic ray spectrum can be found by propagating the sea level muon spectra using Formula \eqref{equ:Surfcosmic}.

\section {CUPP Muon Flux Density as a Function of depth} \label{app:CUPP_depth}
\begin{table}[h!]
\caption{Muon Flux Density as a function of depth measured in CUPP \cite{enqvist2005measurements}.}
\centering
\begin{tabular}{|c|c|}
\hline
Depth        & Flux Density    \\
{[}m.w.e.{]} & {[}$m^{-2}s^{-1}${]}    \\ \hline
0            & $180 \pm 20 $      \\ \hline
210          & $1.3 \pm 0.2 $      \\ \hline
420          & $ (2.3 \pm 0.3) \times 10^{-1}$ \\ \hline
980          & $ (2.1 \pm 0.2) \times 10^{-2}$ \\ \hline
1900         & $ (3.2 \pm 0.3) \times 10^{-3}$ \\ \hline
2810         & $ (6.2 \pm 0.6) \times 10^{-4}$ \\ \hline
3960         & $ (1.1 \pm 0.1) \times 10^{-4}$ \\ \hline
\end{tabular}
\end{table}

\bibliography{Neutron.bib}

\end{document}